\documentclass{PoS}

\title{Strange quark content of the nucleon }

\ShortTitle{Strange quark content of the nucleon }

\author{\speaker{Ronald~Babich}\hspace{-0.3em},$^{\;a}$ Richard~Brower,$^{ab}$
        Michael~Clark,$^b$ George~Fleming,$^c$ James~Osborn,$^d$
	and Claudio~Rebbi$^{ab}$\\ \\
        \llap{$^a$}Department of Physics, Boston University,\\
	590 Commonwealth Avenue, Boston, MA 02215, USA\\
        \llap{$^b$}Center for Computational Science, Boston University,\\
	3 Cummington Street, Boston, MA 02215, USA\\
        \llap{$^c$}Department of Physics, Yale University,\\
	New Haven, CT 06520, USA\\
	\llap{$^d$}Argonne Leadership Computing Facility,\\
	9700 S. Cass Avenue, Argonne, IL 60439, USA\\ \\
        E-mail: \email{rbabich@bu.edu}, \email{brower@bu.edu},
        \email{mikec@bu.edu}, \email{George.Fleming@yale.edu},
        \email{osborn@alcf.anl.gov}, \email{rebbi@bu.edu}}

\abstract{We discuss the calculation of disconnected diagrams needed
  for determining the strange quark content of the nucleon on the
  lattice.  We present results for the strange scalar form factor and
  the related parameter $f_{Ts}$, which enters into the cross-section
  for the scattering of dark matter off nuclei in supersymmetric
  extensions of the standard model.  In addition, we present results
  for the strange contribution to the nucleon's axial and
  electromagnetic form factors.  The calculations were performed with
  two dynamical flavors of Wilson fermions on a $24^3 \times 64$
  anisotropic lattice with $a_s \approx 3 a_t \approx 0.11$~fm and
  $M_\pi \approx 400$~MeV.}

\FullConference{The XXVI International Symposium on Lattice Field Theory \\
		 July 14 - 19, 2008\\
		 Williamsburg, Virginia, USA}

\begin{document}

\section{Introduction}

As the lightest quark not participating in the valence sector, the
strange quark plays a special role in our understanding of the
structure of the nucleon.  In recent years, there has been a vigorous
experimental effort to determine the strange contribution to the
nucleon's elastic electroweak form factors.  In particular, a number
of experiments have sought to measure the strange electric and
magnetic form factors via parity-violating electron
scattering~\cite{Beck:2001yx}, notably SAMPLE, A4, HAPPEX, and G0.
Recent combined analyses~\cite{Liu:2007yi,Young:2006jc} find values
for $G_E^s(Q^2)$ and $G_M^s(Q^2)$ that are small and consistent with
zero in the range of momenta so far explored.

Also of interest is the strange axial form factor $G_A^s(Q^2)$, to
which electron scattering experiments are relatively insensitive.  At
present, the best constraints come from the two-decades old neutrino
scattering data of the E734 experiment at
Brookhaven~\cite{Ahrens:1986xe}.  A recent
analysis~\cite{Pate:2008va}, combining these results with those of
HAPPEX and G0, favors a negative value for $G_A^s(Q^2)$ in the range
$0.45 < Q^2 < 1.0$~GeV$^2$.  In time, a more precise determination should be
possible with results from the new generation of neutrino
scattering experiments (e.g., BooNE, MINER$\nu$A).  A special case is
presented by the strange axial form factor at zero momentum transfer,
$G_A^s(0) = \Delta s$, which may be identified with the strange quark
contribution to the spin of the nucleon.  This quantity is of
particular importance, given the role sea quarks are thought to play
in resolving the proton ``spin crisis.''  In principle, it is
accessible in deep inelastic scattering, where it is given by the
first moment of the helicity-dependent structure function $\Delta
s(x)$.  In practice, however, determining the first moment requires an
extrapolation of the experimental data to small values of $x$, where
uncertainties are less under control.  There is some tension between
the two most recent analyses from
HERMES~\cite{Airapetian:2007mh,Airapetian:2008qf}, which rely on
different techniques; the former favors a negative value for $\Delta
s$ while the latter finds a result consistent with zero, within somewhat
larger uncertainties.

Unlike the strange electromagnetic and axial form factors, the strange
scalar form factor $G_S^s(Q^2)$ is not directly accessible to
experiment.  At zero momentum transfer, this quantity corresponds to
the strange scalar matrix element $\langle N|\bar ss|N\rangle$.  Often
considered in relation to the nucleon-pion sigma term, it is an
important parameter in models of the nucleon.  It has also received
attention, however, for the role it plays in the interpretation of
dark matter experiments.  Many models of TeV-scale physics, including
most supersymmetric extensions of the standard model, yield a dark
matter candidate (e.g., neutralino) that scatters from nuclei via
Higgs exchange.  The Higgs predominantly couples to strange quarks in
the nucleon, and so $\langle N|\bar ss|N\rangle$ enters the
cross-section through the quantity
\begin{equation}
f_{Ts} = \frac{m_s \langle N|\bar ss|N\rangle}{M_N}\,.
\end{equation}
As emphasized recently in~\cite{Bottino:2001dj,Ellis:2005mb}, $f_{Ts}$
is poorly known at present and represents the leading theoretical
uncertainty in the interpretation of direct detection experiments.

Strange form factors have been explored on the lattice in a number of
past investigations, mainly within the quenched approximation.  For a
partial list of references, we direct the reader
to~\cite{Babich:2007jg}.  For some relevant interesting results
presented at this conference, see~\cite{Bali:2008sx,Doi:2008hp,Ohki:2008ge}.

\section{Method}

We begin by defining the basic correlation functions from which
we will extract the strange form factors, starting with the usual two-point
function for the proton.  At momentum $\vec q$, this is given by
\begin{equation}
G^{(2)}(t,t_0;\vec q) = (1+\gamma_4)^{\alpha\beta}
\sum_{\vec x} e^{i\vec q \cdot \vec x}
\langle P^\beta(\vec x,t)\bar{P}^\alpha (\vec 0,t_0)\rangle \,.
\label{eq-g2}
\end{equation}
Here $P^\alpha=\epsilon_{abc} (u_a^T C\gamma_5 d_b) u_c^\alpha$ is the
standard interpolating operator for the proton, with smeared quark
fields, and $(1+\gamma_4)$ projects out the positive-parity state.
Next, we define various three-point functions $G_X^{(3)}(t,t',t_0;q^2)$,
where $X=S,E,A$ correspond to the disconnected scalar, electric, and
axial form factors, respectively.  These are given by
\begin{equation}
G_S^{(3)}(t,t',t_0;\vec q)= (1+\gamma_4)^{\alpha\beta} \sum_{\vec x,\vec x'} 
e^{i\vec q \cdot \vec x'} \langle P^\beta(\vec x,t) [\bar ss(\vec x',t')
- \langle\bar ss(\vec x',t')\rangle] \bar{P}^\alpha (\vec 0,t_0)\rangle 
\label{eq-g3s}
\end{equation}
for the scalar,
\begin{equation}
G_E^{(3)}(t,t',t_0;\vec q)= (1+\gamma_4)^{\alpha\beta} \sum_{\vec x,\vec x'} 
e^{i\vec q \cdot \vec x'} \langle P^\beta(\vec x,t) [V_4(\vec x',t')
- \langle V_4(\vec x',t')\rangle] \bar{P}^\alpha (\vec 0,t_0)\rangle 
\end{equation}
for the electric, and
\begin{equation}
G_A^{(3)}(t,t',t_0;\vec q) = \frac{1}{3}
\sum_{i=1}^3 \sum_{\vec x,\vec x'} e^{i\vec q \cdot \vec x'}
[-i(1+\gamma_4)\gamma_i\gamma_5]^{\alpha\beta}
\langle P^\beta(\vec x,t) [A_i(\vec x',t')-\langle A_i(\vec x',t')\rangle]
\bar{P}^\alpha (\vec 0,t_0)\rangle
\label{eq-g3a}
\end{equation}
for the axial, where $V_\mu$ and $A_\mu$ are the point-split vector
and axial currents.  Note that we always employ the vacuum-subtracted
value of the current, $[J(\vec x,t)-\langle J(\vec x,t)\rangle]$, even
though this is only strictly necessary when $J$ is the scalar density,
since the expectation value of the others vanish.  Given finite
statistics, however, and an inexact estimate of the trace, it is
possible that using the vacuum-subtracted value gives reduced
statistical errors.  Empirically, we find that for the strange scalar
and axial form factors at $q^2=0$, the two approaches give
indistinguishable results.  At larger momenta, however, uncertainties
for the vacuum-subtracted quantities are noticeably smaller.

Inserting a complete set of states in Eq.~(\ref{eq-g2}) yields
the spectral decomposition
\begin{equation}
G^{(2)}(t,t_0;\vec q) = \sum_n c_n(\vec q) e^{-E_n(\vec q)(t-t_0)}\,,
\end{equation}
where the coefficients are given by
\begin{equation}
c_n(\vec q) = 2\left(1+\frac{m_n}{E_n}\right) Z_n^2(\vec q)\,.
\label{eq-cn}
\end{equation}
Here $Z_n(\vec q)$ is the amplitude for annihilating the $n$th state
with our interpolating operator ($n=1$ corresponds to the nucleon
itself).  The momentum dependence in $Z_n$ arises because we utilize
smeared quark sources for the nucleon.  Performing a similar
decomposition for a generic three-point function, we find
\begin{equation}
G_X^{(3)}(t,t',t_0;\vec q) = \sum_{m,n} j_{nm}(\vec q)
e^{-m_n(t-t')} e^{-E_m(\vec q)(t'-t_0)}\,,
\label{eq-g3x}
\end{equation}
where the coefficients $j_{nm}$ may in general be expressed in terms
of some combination of form factors.  In particular, the correlation
functions given in Eqs.~(\ref{eq-g3s}-\ref{eq-g3a}) have been defined
such that a single form factor enters the coefficient $j_{11}$ for
each case, according to
\begin{equation}
j_{11}(\vec q) = 2\left(1+\frac{m_1}{E_1(\vec q)}\right)
Z_1(\vec 0)Z_1(\vec q) G_X^s(q^2)
\end{equation}
for $X=S,E,A$, where $G_X^s(q^2)$ is the corresponding strange form factor
of the nucleon.  In terms of the coefficients $c_n$ extracted from the
two-point function $G^{(2)}(t,t_0;\vec q)$, this becomes
\begin{equation}
j_{11}(\vec q)=G_X^s(q^2) \sqrt{\frac{1}{2}\left(1+\frac{m_1}{E_1(\vec q)}
\right) c_1(\vec 0) c_1(\vec q)} \,.
\label{eq-j11}
\end{equation}
Our general strategy will be to fit the correlation functions
$G_X^{(3)}(t,t',t_0;\vec q)$ to Eq.~(\ref{eq-g3x}), taking into
account both the ground state nucleon and a single excited state.  We
may then extract the nucleon form factors from $j_{11}$ with input
from the two-point function.  In principle, one could also obtain form
factors of the excited state from $j_{22}$, as well as transition form
factors from $j_{12}$ and $j_{21}$.  In practice, however, we expect
these to absorb the contributions of still higher states and trust
only the ground state form factors to be reliable.

The motivation for this approach will be discussed further as we
present our results in the following section.  Before turning there,
however, we conclude this section with a few remarks on the
calculation of the disconnected insertion.  This represents a major
challenge for the lattice since it requires the trace of the full
Dirac propagator over color, spin, and spatial indices.  Our approach
for estimating this trace is described in~\cite{Babich:2007jg}.
Briefly, it involves inverting against a set of sources where each is
nonzero on only a small number of sites, with these sites as widely
separated as possible.  In the parlance of the usual stochastic source
method, this is equivalent to using a single noisy source with
``extreme dilution.''  It provides a very accurate estimate of the
trace, and since the small amount of contamination arising from
off-diagonal terms is not gauge-invariant, the method is unbiased.  On
our $24^3\times 64$ lattice, we have chosen a dilution pattern where
the smallest spatial separation between sites is $6\sqrt{3}a_s$.  With
864 (times 12 for color/spin) such sources per lattice, we obtain the
trace over each of four time-slices.

In order to construct the nucleon correlators, we also calculate 64
propagators at the light quark mass per lattice, with one originating
from each time-slice.  We utilize gaussian smearing for the light
quark propagators.

\section{Preliminary results}

The results presented here were calculated on a $24^3\times 64$
anisotropic lattice, using an ensemble of 863 gauge configurations
provided by the Hadron Spectrum Collaboration.  These were generated
with two dynamical flavors of unimproved Wilson fermions, with
$M_\pi\approx 400$~MeV and $a_s = 0.108(7)\,\mathrm{fm} \approx 3a_t$.
Given the anisotropy, this lattice has a relatively short extent in
time, which has influenced our choice of method.  Conventionally, one
extracts the form factors by considering various ratios of the three-
and two-point functions defined above.  For example, at zero momentum
transfer, one has
\begin{equation}
R_X(t,t',t_0;q^2=0) = \frac{G_X^{(3)}(t,t',t_0;\vec 0)}
{G^{(2)}(t,t_0;\vec 0)} \rightarrow G_X^s(q^2)\,,
\end{equation}
for large time separations.  As discussed in the previous section, we
have chosen instead to fit the three-point function that appears in
the numerator of this ratio directly.  The reasons are two-fold.
First, this allows us to avoid contamination from backward-propagating
states, which are problematic due to the short temporal extent of our
lattice.  At the same time, it allows us to explicitly take into
account the contribution of (forward-propagating) excited states.

\begin{figure}
\begin{center}
\includegraphics*[width=0.6\textwidth]{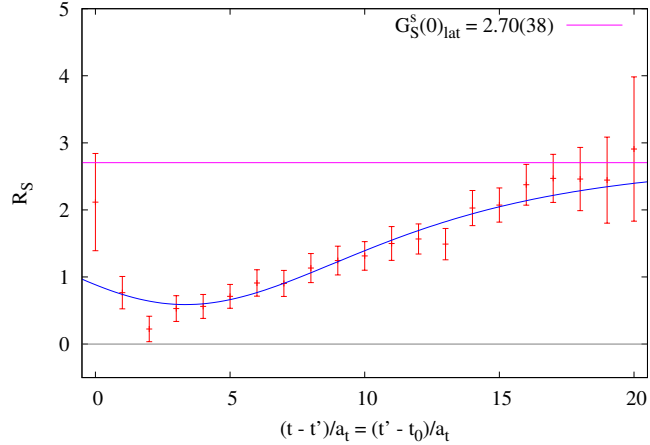}
\caption{\label{fig-scalar-q0}Results for the scalar form factor at $q^2=0$.}
\end{center}
\end{figure}

Our fit of the scalar form factor at $q^2=0$ is shown in
Figure~\ref{fig-scalar-q0}.  For the purpose of plotting, we have
normalized our results by a fit to the two-point function.  With this
normalization, dominance of the ground state should manifest as a
plateau at large times.  The statistical error in the extracted form
factor, $G_S^s(0)_\mathrm{lat} = 2.70(38)$ is determined by jackknife
over the full fitting procedure.  We note that in this preliminary
analysis, we have treated the system symmetrically by requiring that
the separation between the source and current insertion equal that
between the insertion and the sink, i.e., $(t-t')=(t'-t_0)$.  It
should be possible to reduce the statistical errors further by lifting
this constraint.  We expect that this will prove especially
advantageous for the results at nonzero momentum transfer presented
below, since these require a fit to a sum of four, rather than three,
exponentials.

We have not yet attempted to quantify the systematic uncertainties.
In addition to the usual systematics associated with the finite
lattice spacing, finite volume, and the unphysically heavy mass of the
light quarks, our approach is also sensitive to the choice of
the fitting window in the fit of the two-point function.  Recall that
in order to extract the form factors from $G_X^{(3)}(t,t',t_0;\vec
q)$, we must first determine the coefficients $c_n(\vec q)$ and
masses/energies $E_n(\vec q)$ from a fit to $G^{(2)}(t,t_0;\vec q)$.
Since we have access to a total of $863 \times 64 = 55,232$ nucleon
correlators, these tend to be very well-determined.  The coefficients
$c_n$ are somewhat sensitive to the choice of fitting window, however,
and since they multiply the form factor in Eq.~(\ref{eq-j11}), this
translates into a direct systematic error on the form factor, at
perhaps the ten percent level.

Given our determination of the strange scalar matrix element, for the
renormalization-invariant quantity $f_{Ts}$, we estimate
\begin{equation}
f_{Ts} = \frac{m_s \langle N|\bar s s|N\rangle}{M_N} = 0.51(8)(3)\,,
\label{eq-fts}
\end{equation}
where we have inserted the physical nucleon mass for $M_N$.  The
second error is the uncertainty in relating this mass to the lattice
scale, the first is statistical, and no other systematics have been
taken into account.  The bare quark mass $m_s$ is the naive mass
appearing in the Wilson action, minus the critical mass, as determined
from a (partially quenched) chiral extrapolation of the mass of the
pseudoscalar.  It is worth noting that the matrix element appearing in
Eq.~(\ref{eq-fts}) was calculated for a world with a 400~MeV pion.
This is a consistent determination of the physical value only insofar
as the matrix element $\langle N|\bar ss|N\rangle$ is insensitive to
the mass of the light quarks.  As an alternative, we may work
consistently in such a world by inserting our calculated value of the
nucleon mass.  In this case, the scale dependence drops out, and we
find $f_{Ts}=0.44(7)$.

\begin{figure}
\begin{minipage}[t]{0.48\textwidth}
\includegraphics*[width=\textwidth]{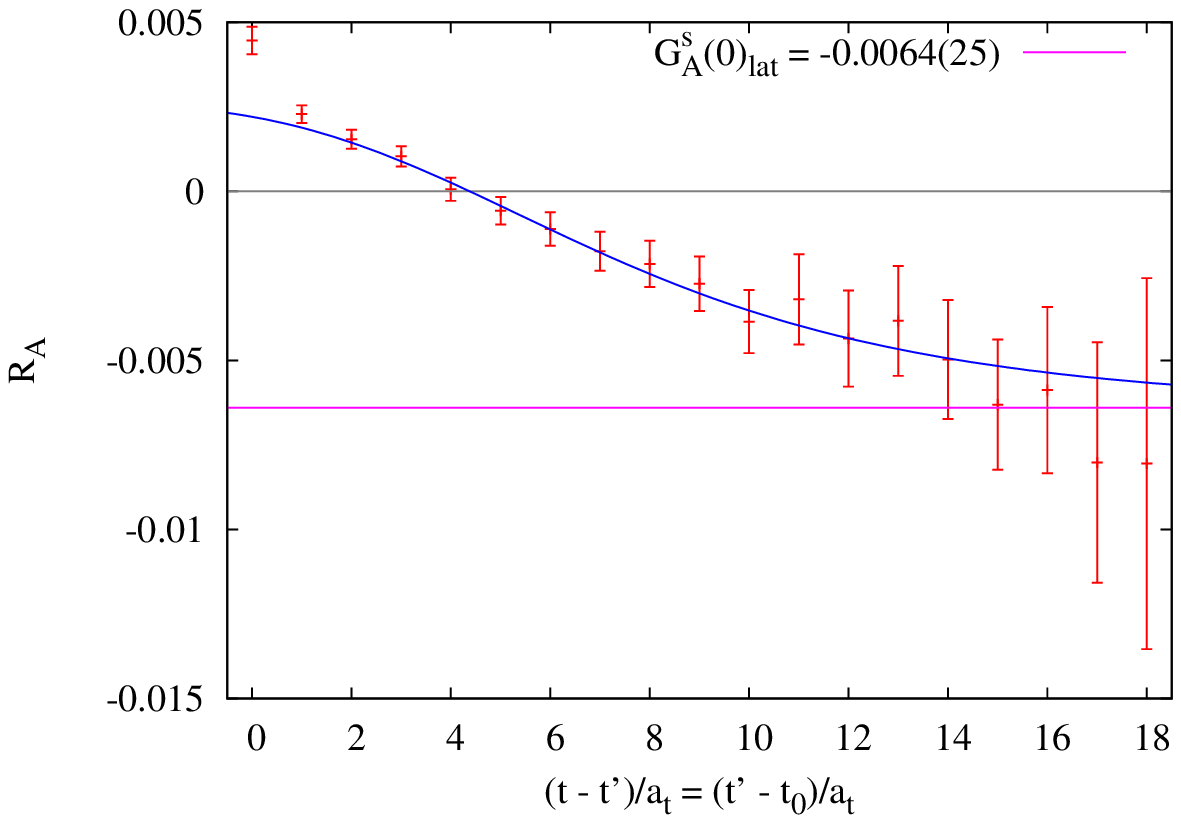}
\caption{\label{fig-axial-q0}Results for the strange axial form factor
  at $q^2=0$.}
\end{minipage}
\hfill
\begin{minipage}[t]{0.48\textwidth}
\includegraphics*[width=\textwidth]{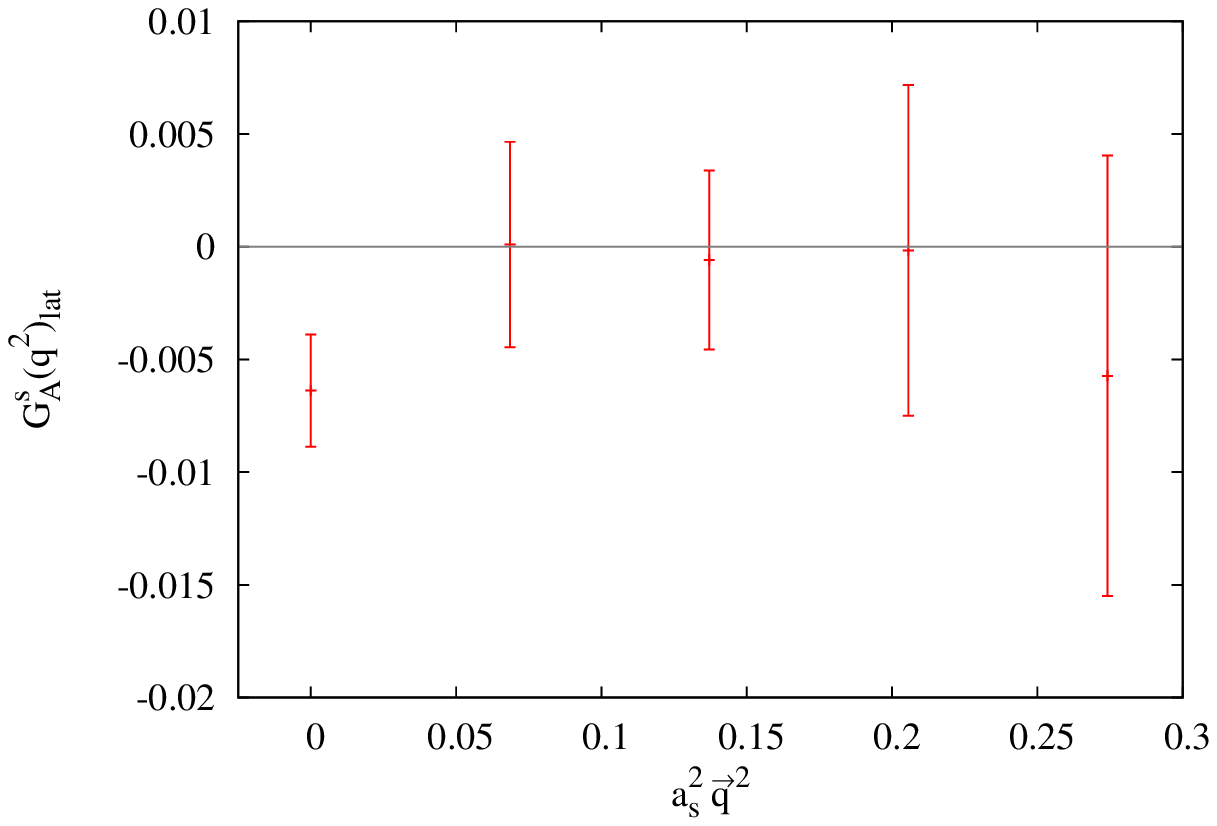}
\caption{\label{fig-axial}Strange axial form factor as function of
  momentum.}
\end{minipage}
\end{figure}

Results for the strange axial form factor are shown in
Figure~\ref{fig-axial-q0}.  As for $G_S^s(0)$, we note that our result
$\Delta s = G_A^s(0) = -0.064(25)$ has not been renormalized and so
may not be compared to the experimental results.  It is noteworthy,
however, that we find a value that is negative and distinct from zero
at the level of 2.5$\sigma$.  In Figure~\ref{fig-axial}, we show the
momentum dependence of $G_A^s(q^2)$.

\begin{figure}
\begin{minipage}[t]{0.48\textwidth}
\includegraphics*[width=\textwidth]{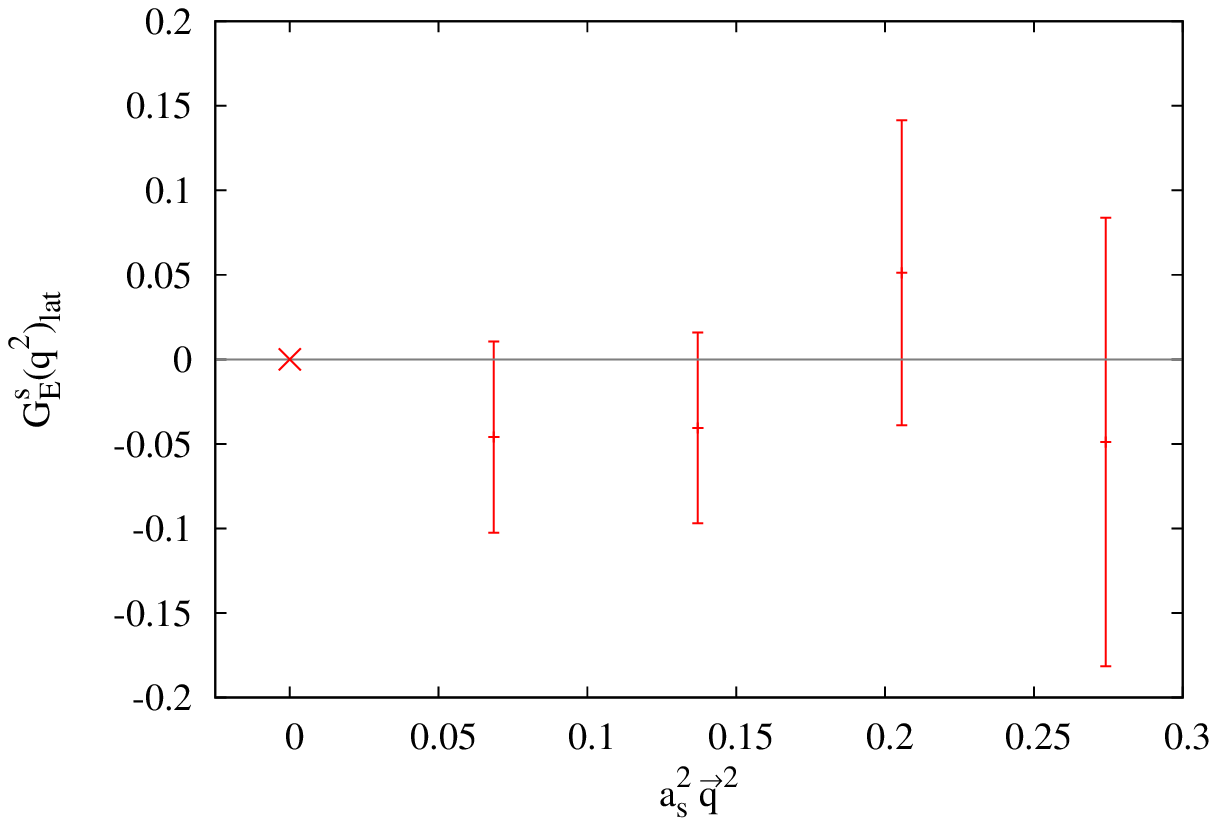}
\caption{\label{fig-electric}Strange electric form factor as function
  of momentum.}
\end{minipage}
\hfill
\begin{minipage}[t]{0.48\textwidth}
\includegraphics*[width=\textwidth]{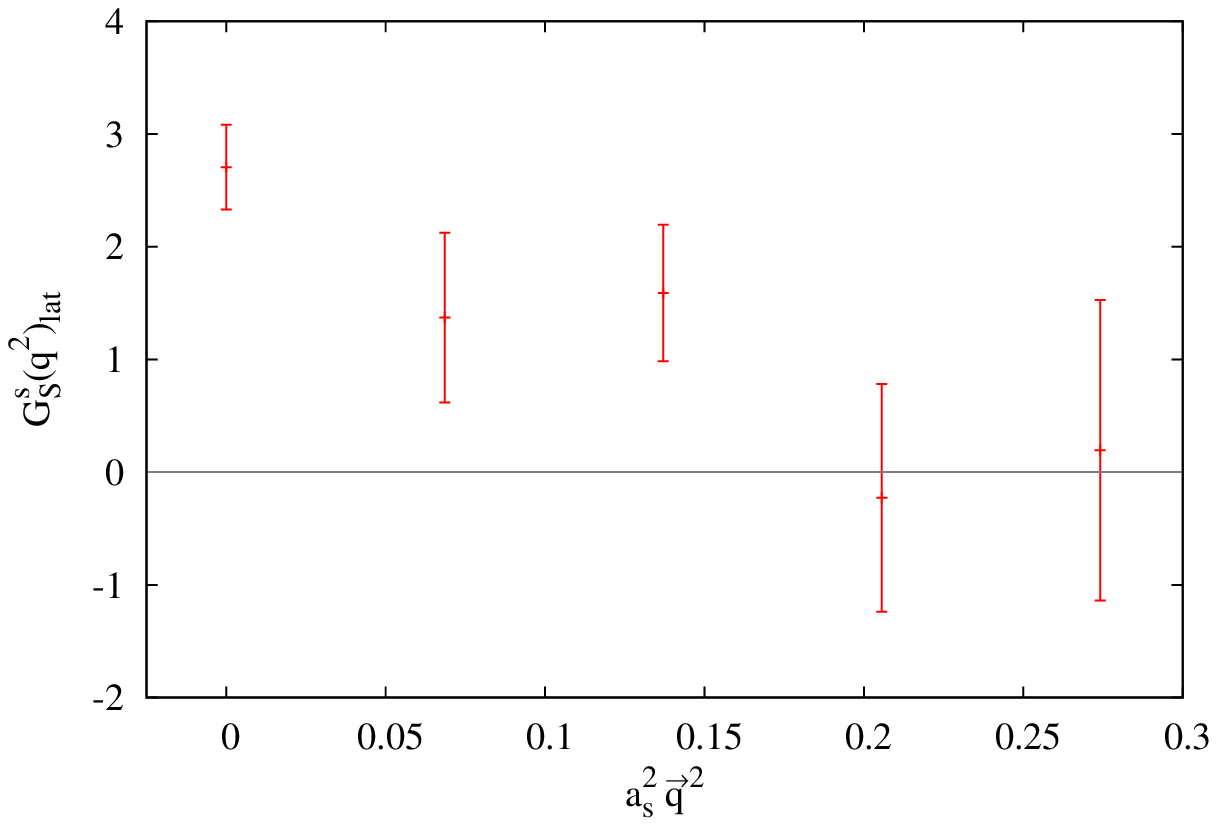}
\caption{\label{fig-scalar}Strange scalar form factor as function of
  momentum.}
\end{minipage}
\end{figure}

Finally, in Figure~\ref{fig-electric}, we present our preliminary
results for the strange electric form factor as a function of
momentum.  Note that since the strange quark does not contribute to
the electric charge of the nucleon, $G_E^s(0)$ must vanish.  For
completeness, the momentum dependence of the strange scalar form
factor is shown in Figure~\ref{fig-scalar}.

\acknowledgments

This work was supported in part by US DOE grants DE-FG02-91ER40676 and
DE-FC02-06ER41440, NSF grants DGE-0221680 and PHY-0427646, and by the
National Science Foundation through TeraGrid resources provided by the
Texas Advanced Computing Center~\cite{catlett2007tao}.  We also thank
Boston University and Jefferson Lab for use of their scientific
computing facilities.

\bibliographystyle{JHEPcaps}
\bibliography{lat08_babich}

\end{document}